\documentclass[fleqn,twoside]{article}
\usepackage{espcrc2}
\usepackage{graphicx}
\usepackage{here}

\newcommand{\AmS}{{\protect\the\textfont2
    A\kern-.1667em\lower.5ex\hbox{M}\kern-.125emS}}

\def\beq{\vspace*{-0.2cm}\begin{equation}}
\def\eeq{\end{equation}\vspace*{-0.2cm}}

\def\frac#1#2{{#1\over#2}}
\def\ftild{{\tilde f}}
\def\ftildRe{ \Re e \ftild}
\def\ftildIm{ \Im m \ftild}
\def\therho{\theta\rho}

\title{ Low-mass scalar production in $\gamma \gamma$ scattering}
\author{G. Mennessier
        \address {\footnotesize L.P.T.A., UMR 5207, Universite Montpellier II, CNRS}
      }
\begin{document}

\font\myRm=cmr12  scaled \magstep1
\def\myBf{\bf}

\begin{abstract}
\noindent
  We estimate the $I=0$ scalar meson $\sigma/f_0(600)$ $\gamma\gamma$ widths, 
from $\pi\pi$ and $\gamma\gamma$ scattering data below 700 MeV
using an improved analytic K-matrix model.
\end{abstract}

\maketitle

Communication on part of a work done with

{NARISON Stephan and OCHS Wolgang \cite{ME_NA_O,ME_MI_NA_O}

\section{Introduction. Preliminary remarks}

- This is an attempt to get information on the nature of the low-mass scalar meson,
the controversial  {\myBf $\sigma $} or {\myBf $ \epsilon $} or {\myBf $f_0(600) $},
from  {\myBf ${\gamma + \gamma  \rightarrow \pi + \pi }$ } at energies below
$\sim 700 \  {\rm Mev}$.

- Studies of this process go back to Lyth (1971) \cite{LYTH},
Yndurain (1972) \cite{YNDUR1},
... \cite{BAB_BAS_CA_GOU_ME} \cite{MEN_1} \cite{MOR_PENN_1},
and recently Boglione-Pennington \cite{BOG_PENN_1}, Pennington \cite{PENN1},
Achasov-Shestakov \cite{ACHA_SHES_1},
Oller-Roca \cite{OLL_RO_1}, Pennington et al. \cite{PENN2},
Giacosa-Gutsche-Lyubovitskij \cite{GIA_GUT_LYU} ...

- Though not proved for composite particules (to my knowledge),
  I will assume "usual" analyticity properties in $s$ (the energy squarred),
  with {\myBf Left} cut from $t, u$ particle exchanges,
  and {\myBf Right} cut above threshold, from physical channel.

- Working at lowest order in E.M. , {\myBf unitarity is linear}, and involves
STRONG amplitudes.

- from unitarity and analyticity, Muskhelishvili \cite{MUSKH} have shown how to
determine a set of {\myBf Fundamental solutions}, from which one can obtain
the full family of solutions, which is determined {\myBf up to Polynomial Ambiguity}
once we are given the Left singularities.

- In the 1-channel purely elastic case, the Fundamental solution reduces to the
Omnes formula \cite{OMNES}, where the phase shift must be chosen  {\myBf continuous},
to get an analytic and invertible function.

- Analytic extrapolation is unstable if there are no bounds in all directions
of the complex plane, and sensitive to even small but rapid variations.
This is, presumably, one of the reason of the controverses
on the low-mass scalar meson, which appear to be a very broad object.
See a recent discussion by Yndurain et al. \cite{YND_GAR_PEL}

\section{STRONG interaction parametrisation}

  Since we only study the low energy part, we assume {\myBf elastic unitarity}
from threshold up to infinity, and unitarise only the S-waves.

Here we will neglect, in particular, the opening of the $K-\bar K$ threshold,
the effect of the f2(1270) ...

  We will use parametrisations, of generalized {\myBf analytic K-matrix} \cite{BAB1} type,
which allow explicit expressions for the Fundamental solution,
and explicit continuation on the Riemann sheets.

\subsection{I=0, S-wave}
-  We use an ${ \cal N / \cal D}$ representation for $T^{0}$ :
\beq
  T^{0} (s) = \frac{G  f_0(s)} {s_R - s - G \ftild_0 (s) }
\eeq
\beq
  {\cal D} = s_R - s - G \ftild_0 (s);
\eeq
with Fundamental solution
\beq
  F^0 (s) = 1/{\cal D};
\eeq
where the {\it shape function} $f_0(s)$ has only {\myBf Left} singularities,
while $\ftild_0 (s)$ has only {\myBf Right} singularities, with
\beq
  {\ftildIm}_0(s) = \therho (s) f_0(s)
\eeq
and $\ftildRe_0 $ is obtained by dispersion relation, with minimal subtraction at $s=0$.
For $G$ small, there would exist a {\it bare} pole at $s = s_R$.

\subsubsection{ }
We choose a simple form for $f_0(s)$,
\beq
  f_0(s) = \frac{ s - s_{A0} }{ s + \sigma _{D0} }
  \label{f0pol}
\eeq
which have an {\myBf Adler zero} and 1 pole to simulate the near Left singularities.

- The I=0, S=0 phase-shifts $\delta^0_0$ have been determined by several groups,
in particular using ROY equations \cite{CCGL1,KA_PE_YND_1}.

To determine the 4 parameters ($ s_{A0}, \sigma _{D0}, s_R , G $),
we fit the phase-shifts $\delta^0_0$ below $800 Mev$,
obtained by Caprini et al. \cite{CCGL1}.

For 26 points, total $\chi^2 = 0.55$, one obtains
$ s_{A0}=0.0167 Gev^2, \sigma _{D0}=0.5013 Gev^2, s_R=0.8232 Gev^2, G=1.1839 $.

The $T^{0}$ amplitude has 2 poles in the second sheet, P1 and P2, with energy w
and energy squarred s values

$ wP1 =0.422 - i \ 0.290 \ GeV; sP1 =0.0936 - i \ 0.2447 \ GeV^2 $

$ wP2 =1.043 - i \ 0.672 \ GeV; sP2 =0.6360 - i \ 1.4027 \ GeV^2 $

  The first one, P1, is not far from the \cite{CCGL1} one : $ 0.441 - i \ 0.272 \ GeV $.
  The second, P2, unexpected for the author, will certainly move a lot when taking
into account what happens near {$K\bar K$} threshold.

  However, if one just take the limit $G \rightarrow 0$ ,
the heavy pole P2 goes to the {\it bare} pole ($ wP2 -> \sqrt s_R $) ,
while P1 goes to unphysical negative s value, on the Left cut
($ wP1 -> \sqrt (- \sigma _{D0}) $).

\subsubsection{ }
For a choice of $ f_0(s) $ with a cut instead of a pole,
\beq
  f_0(s) = \lambda + G \ \frac{1}{s - sD} \ log(1 + \frac{s - sD}{mu2} )
\eeq
and corresponding $\ftild_0 (s)$,
fitting again the phase-shifts \cite{CCGL1}, one obtains with $\chi^2 =0.66$
$ sD= 0.0778,  mu2= 0.3462, G= -0.4673, \lambda= 1.477 $.

When scaling  both $\lambda \ and \ G$ to zero,
the lowest pole P1 disappear before reaching Left cut (and before couplings vanish),
while P2 keep the same behaviour,
going to the {\it bare} pole ($ wP2 -> \sqrt {s_R} $).

  Though this does not correspond to a true QCD limit, it could indicate that the
existence of the P1 pole is related to possibility of physical decay.

\subsection{I=2, S-wave}
- We take for $T^{2}$ :
\beq
  T^{2} (s) = \frac{\Lambda  f_2(s)} {1 - \Lambda \ftild_2 (s) }
\eeq
where
\beq
  f_2(s) = \frac{ s - s_{A2} }{(s + \sigma _{D1})(s + \sigma _{D2})}
\eeq
is more convergent to avoid an unwanted bound state pole,
with
\beq
  {\ftildIm}_2(s) = \therho (s) f_2(s)
\eeq

\section{EM interaction}

\subsection{Pion exchange}
The charged $\pi^{+} \pi^{-}$ production in $\gamma \gamma$ scattering,
is dominated by the Pion exchange,
but which does not contribute to the $\pi^{0} \pi^{0}$. However once produced,
the charged pions can rescatter also into neutral ones.

Let $ \alpha f^B $ be the S-wave projection of total Born Pion exchange.
Then
\beq
  T_{\gamma}^0 = \sqrt{2/3}\ \alpha f^B \ and \ T_{\gamma}^2 = \sqrt{1/3}\ \alpha f^B
\eeq
are the corresponding isospin I=0 and 2 amplitudes.

\subsubsection{}
Let us define  $\ftild^B_0$, analytic on the Left, with
\beq
  {\ftildIm}^B_0 = \therho (s) \ f_0(s) \ f^B(s)
\eeq
{\myBf subtracted at s= 0 to satisfy Thompson limit }, in a minimal way.

Then
\beq
  T_{\gamma}^0 =  \sqrt{2/3} \alpha (f^B + G \frac{\ftild^B_0}{\cal D})
                  + \alpha P(s) F^0
\eeq
is analytic, unitary for any polynomial $P(s)$.
The $\ftild^B_0$ term is naturally interpreted as the rescattering contribution,
and the $P(s) F^0$ as a direct one.
This interpretation is not completely unambiguous, since over-subtracting
(or renormalising differently the unitarisation s-bubbles)
can lead to a different definition, with same total amplitude.
Here also, not to violate Thompson limit, one must restrict to $P(s)$
vanishing at $s=0$,
and limit its degree to avoid too divergent partial-wave.

Thus we will choose $P(s) = \sqrt{2} F_{\gamma} \ s$.
\begin{figure}[h]
\includegraphics[width=9.cm]{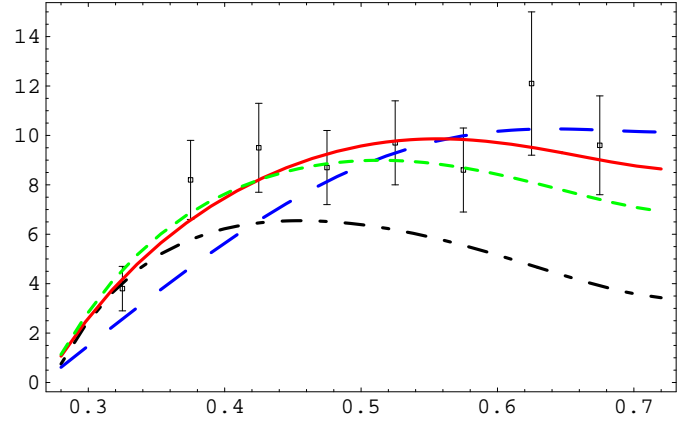}
\caption{\footnotesize
  Fit of the $\pi^0\pi^0$ cross-section  in nanobarn (nb) versus $\sqrt{s}$
using unitarized Born amplitude: $F_\gamma=0$ (dot-dashed);
$F_\gamma=-0.09$: I=0 (large dashed), I=0+2 (continuous);
$F_\gamma=-0.07$: I=0+2 (small dashed).
The data are from Crystal Ball \cite{CRYSTAL} for $|\cos\theta| \leq 0.8$;
}
\label{fig:fitneutral}
\end{figure}
\subsubsection{}
  Analogously, for I=2,
one defines $\ftild^B_2$, analytic on the Left, with
\beq
  {\ftildIm}^B_2 = \therho (s) \ f_2(s) \ f^B(s)
\eeq

Then
\beq
  T_{\gamma}^2 = \sqrt{1/3}\ \alpha
                (f^B + \Lambda \frac{\ftild^B_2} {1 - \Lambda \ftild_2 (s)})
\eeq
is analytic, unitary. Limiting the degree of possible polynomial to the behaviour
of rescattering terms completely suppress polynomial contribution.

  There is then only 1 parameter free, $F_{\gamma}$.

\begin{figure}[hbt]
\includegraphics[width=9.cm]{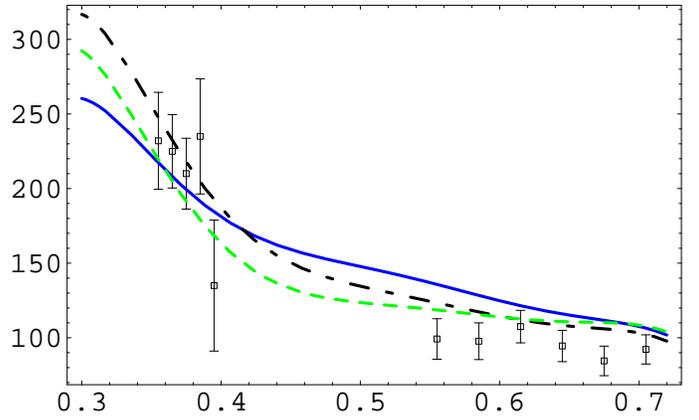}
\caption{\footnotesize 
  The same as in Fig. \ref{fig:fitneutral} but for $\pi^+\pi^-$
using unitarized Born amplitude: $F_\gamma=0$ (dot-dashed);
$F_\gamma=-0.07$ and I=0+2 (small dashed).
The continuous line corresponds to the non-unitarized Born amplitude with $F_\gamma=0.$
The data are from MARKII \cite{MARK2} for $|\cos\theta| \leq 0.6$.
}
\label{fig:fitcharged}
\end{figure}
\subsection{Results}
One uses the $f_0$ in Eq(\ref{f0pol}).
Fit to  MARK II data \cite{MARK2} for $\pi^{+} \pi^{-}$,
and to CRYSTAL BALL \cite{CRYSTAL} for $\pi^{0} \pi^{0}$ below $0.7 Gev$
gives $F_{\gamma} \sim  -0.08  $

This corresponds to residues at the P1 pole
  $resc =( 0.091, 0.116)$ for rescattering,
  $direct=(0.007, 0.031)$ for the direct contribution,
  $tot=( 0.098, 0.151)$ for the total photon-photon width,
which can be translated into (using full hadronic width=580.0 MeV)
into partial widths
$\Gamma$ resc= 2.805, $\Gamma$ direct= 0.126, $\Gamma$ tot= 4.0 keV .

Stephan Narison will speak on consequences for the nature of the particles
associated to the poles.

\end{document}